\documentclass[11pt]{article}
\usepackage[T1]{fontenc}
\usepackage{lmodern}
\usepackage[margin=0.8in]{geometry}
\usepackage{amsmath,amssymb,booktabs,longtable,array,graphicx}
\usepackage[numbers,sort&compress]{natbib}
\usepackage{microtype}
\usepackage{placeins}
\usepackage[hidelinks]{hyperref}
\hypersetup{
  pdftitle={Revisiting the ABCs of Working with AI: Evidence from Radiologists},
  pdfauthor={Daniel Martin}
}

\newcommand{\Description}[1]{}
\newcommand{\MainAbilityEstimate}{-0.685}
\newcommand{\MainAbilitySE}{0.247}
\newcommand{\MainCalibrationEstimate}{0.493}
\newcommand{\MainCalibrationSE}{0.221}
\newcommand{\MainAbilityP}{0.005}
\newcommand{\MainCalibrationP}{0.026}
\newcommand{\MainUnassistedErrorRate}{9.91}
\newcommand{\MainAssistedErrorRate}{9.39}
\newcommand{\MainPairedUnassistedErrorRate}{10.03}
\newcommand{\MainDifferenceSD}{9.380}
\newcommand{\AccuracyOnlyEstimate}{-0.339}
\newcommand{\AccuracyOnlySE}{0.174}
\newcommand{\AccuracyOnlyP}{0.052}
\newcommand{\StandardizedOutcomeAbilityEstimate}{-0.073}
\newcommand{\StandardizedOutcomeAbilitySE}{0.026}
\newcommand{\StandardizedOutcomeCalibrationEstimate}{0.053}
\newcommand{\StandardizedOutcomeCalibrationSE}{0.024}
\newcommand{\TwoWayAbilitySE}{0.246}
\newcommand{\BootstrapAbilityLow}{-1.158}
\newcommand{\BootstrapAbilityHigh}{-0.185}
\newcommand{\TieAbilityChange}{-0.001551}
\newcommand{\TwoWayCalibrationSE}{0.217}
\newcommand{\BootstrapCalibrationLow}{0.065}
\newcommand{\BootstrapCalibrationHigh}{0.962}
\newcommand{\TieCalibrationChange}{0.000006}
\newcommand{\BootstrapReps}{999}
\newcommand{\BootstrapSeed}{20260728}
\newcommand{\ReaderProbabilityTies}{2}
\newcommand{\ReferenceProbabilityTies}{0}

\title{Revisiting the ABCs of Working with AI: Evidence from Radiologists}
\author{Daniel Martin\\[3pt]
  \small Department of Economics\\
  \small University of California, Santa Barbara\\
  \small \href{mailto:danielmartin@ucsb.edu}{danielmartin@ucsb.edu}\\
  \small ORCID: \href{https://orcid.org/0000-0001-6483-3923}{0000-0001-6483-3923}}
\date{November 15, 2026}
\begin{document}
\maketitle
\begin{abstract}
I examine the relationship between the benefits of working with AI and two baseline traits --- ability and belief calibration --- among all 227 radiologists available in
Collab-CXR, a dataset from an experiment where AI assistance was provided
during X-ray diagnosis. When analyzed jointly, both lower unassisted accuracy
and smaller miscalibration, measured by the absolute gap between average
confidence and accuracy, are associated with greater improvements in
accuracy with AI. In addition, the association
between unassisted accuracy and improvement with AI becomes
more negative and statistically significant when controlling for calibration.
\end{abstract}

\section{Introduction}

Who benefits from working with AI? A user who makes more mistakes without
assistance has more room to improve, but using advice also requires
judging when to rely on it. Ability and confidence in one's judgments can
therefore have distinct roles. Caplin et al.~\cite{caplin2026abcs} organize
these characteristics as the ``ABCs'' of working with AI: ability, beliefs,
and calibration. In their image-classification experiment, Prolific
participants with lower ability benefit more from AI and, holding ability
constant, participants whose average confidence is closer to their
accuracy benefit more.

I test whether these associations extend to experts making diagnostic
judgments. I use Collab-CXR~\cite{moehring2025dataset,moehring2025repository},
public data from the radiologist experiment of
Agarwal et al.~\cite{agarwal2023radiology}, in which radiologists assess
retrospective chest X-ray cases with and without AI predictions.

In Caplin et al.'s experiment, participants report the probability that a person in a photograph
is over 21 years old. These reports determine both the implied binary
classification and confidence in that classification. Ability is measured with average
accuracy, and calibration is measured with the negative absolute gap between average
confidence and accuracy. I apply this construction to radiologists'
probabilities of diagnostic findings in the Collab-CXR dataset. 

Yu et al.~\cite{yu2024heterogeneity} use this dataset to evaluate unassisted performance using
absolute probability error and the area under the receiver operating
characteristic curve (AUROC) among 140 radiologists from two designs.
They find that unassisted performance does not consistently predict AI
benefits. Similarly, when I analyze -- among all 227 radiologists across three
designs -- binary accuracy alone, it has a negative association with improvement that is
not significant at the 5\% level ($p=\AccuracyOnlyP$). 

However, when I instead jointly examine binary accuracy and average confidence
calibration, the accuracy coefficient becomes more negative and both
associations are significant at 5\%. Analyzing them jointly shows that, among radiologists
with similar accuracy, a smaller absolute confidence gap is associated with greater
improvement in accuracy with AI.

\section{Data and measurement}

Each radiologist reviews patient cases and assesses multiple diagnostic
findings on each chest X-ray. After
excluding warm-up rounds and pilot participants and retaining the 15 tasks
with AI predictions, there are 314,400 recorded diagnostic probabilities
from 227 radiologists reviewing 324 patient cases. I aggregate these
assessments into 8,500 regression observations, one for each
radiologist--patient pair.
Supplementary Section~\ref{supp-sec:sample} lists the tasks and exclusions.

Design~1 contributes 159 radiologists, Design~2 contributes 33, and Design~3
contributes 35.
The three designs differ in whether assistance is evaluated on the same
patient cases~\cite{moehring2025dataset}. In Design~1, each radiologist assesses separate cases with
and without AI. Design~2 observes each case under all four combinations of AI
assistance and clinical history, with separated sessions and varied
information sequences. Design~3 introduces assistance during a second
review of the same 50 cases, holding history status fixed, so its
comparison combines AI access with repeat exposure. I pool clinical-history
conditions, giving them equal weight when both are observed for a
radiologist, patient, and AI state.
The outcome therefore compares assisted and unassisted accuracy for the
same patient in Designs~2 and~3. In Design~1, it compares accuracy for a
case assessed without AI with the radiologist's average assisted accuracy on
separate cases.

I measure accuracy by agreement with a reference formed from the mean
probability assigned by five expert radiologists. For a given assessment,
let $b\in[0,1]$ be the recorded radiologist probability and $q\in[0,1]$
the expert reference. Define a binary decision
$d=\mathbf{1}\{b>0.5\}$, reference label $s=\mathbf{1}\{q>0.5\}$,
agreement $a=\mathbf{1}\{d=s\}$, and decision confidence
$c=\max\{b,1-b\}$. Confidence is the probability assigned to the chosen
binary diagnosis: $b$ for a positive decision and $1-b$ for a negative
decision.

\section{Empirical strategy}

Using the same radiologist assessment without AI assistance to measure both
baseline ability and improvement with AI would make an unusually poor
assessment both lower measured ability and increase measured improvement.
Caplin et al.'s experiment overcomes this
challenge by separating the images used to measure traits from those used
to measure outcomes. The experiment here has fewer distinct cases per
participant, so instead I adapt the split-sampling logic of
Yu et al.~\cite{yu2024heterogeneity} by constructing unassisted traits
separately for each patient case, excluding that case's assessments. This removes the direct
overlap between the outcome and the assessments used to construct the traits.
For radiologist $h$ and patient case $i$, let $\bar a_{h,-i}(0)$ denote average
unassisted accuracy and $\bar c_{h,-i}(0)$ denote average unassisted confidence,
both calculated using the radiologist's assessments of all patient cases
other than $i$. These averages use the recorded assessments across diagnostic findings and available
clinical-history conditions. Define
\[
 A_{h,-i}=\bar a_{h,-i}(0),\qquad
 C_{h,-i}=-\left|\bar c_{h,-i}(0)-\bar a_{h,-i}(0)\right|.
\]
I refer to $A_{h,-i}$, the unassisted accuracy on other patient cases, as
baseline \emph{ability}, and $C_{h,-i}$, the negative absolute difference
between this accuracy and average unassisted decision confidence, as
baseline \emph{calibration}.

Let $Y_{hi}(t)$ denote radiologist $h$'s accuracy on patient case $i$ in AI
state $t$, where $t=0$ denotes no AI and $t=1$ denotes assistance,
averaging the recorded assessments across diagnostic findings and available
clinical-history conditions. I measure
improvement in accuracy with AI as accuracy with AI minus accuracy
without AI, in percentage points. Negative values indicate a decline.
For Designs~2 and~3, the outcome is
$g_{hi}=100[Y_{hi}(1)-Y_{hi}(0)]$.
For Design~1, it is
$g_{hi}=100[\bar Y_h(1)-Y_{hi}(0)]$, where $\bar Y_h(1)$ averages
the separate assisted cases. Thus each design contributes one regression
observation per radiologist-patient pair. In Design~1, patient case $i$
is assessed without AI.

I estimate
\[
 g_{hi}=\beta_0+\beta_A A^z_{h,-i}+\beta_C C^z_{h,-i}
       +\gamma_2\mathbf{1}\{D_h=2\}
       +\gamma_3\mathbf{1}\{D_h=3\}+\varepsilon_{hi},
\]
where $D_h$ is the experimental design and the superscript $z$ denotes
standardization. The ability and calibration coefficients describe the
association between a one-standard-deviation increase in the corresponding
unassisted trait and improvement in accuracy with AI, in percentage points,
holding the other regressors fixed.
These coefficients describe associations, not the effects of changing a
radiologist's accuracy or calibration.

To standardize each predictor, I first average its values across patient cases
within radiologist, then use the unweighted mean and sample standard
deviation across radiologists to scale every observation. The primary
regression weights radiologist-patient observations equally.
Radiologists contribute 30, 60, and 50 observations in Designs~1, 2,
and~3, respectively. In the supplement, I examine an alternative that gives
each radiologist equal total regression weight.

\section{Results}

\begin{figure}[!htbp]
\centering
\includegraphics[width=\linewidth]{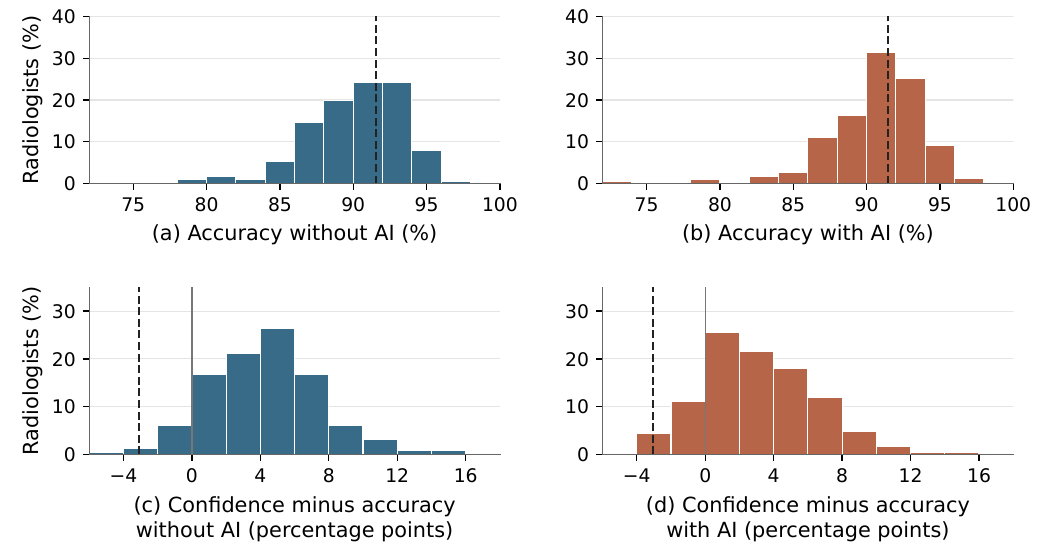}
\caption{Distributions of accuracy and net confidence, following Figure~1 of
Caplin et al.~\cite{caplin2026abcs}. Each of the 227 radiologists contributes
one mean per condition, using all retained assessments. Net confidence is
mean decision confidence minus accuracy. Bins span two percentage points.
Dashed lines show mean AI-only scores on the corresponding case sets,
weighting radiologists equally.}
\Description{Four histograms show radiologist accuracy and confidence minus
accuracy, with and without AI. Vertical dashed lines mark the corresponding
AI-only benchmarks.}
\label{fig:distributions}
\end{figure}

Across the 8,500 regression observations, mean unassisted accuracy is
90.09\% and mean assisted accuracy is 90.61\%, corresponding to error rates
of \MainUnassistedErrorRate\% and \MainAssistedErrorRate\%. Average accuracy
is therefore 0.52 percentage points higher with AI. With equal radiologist
weights, the accuracy means are 90.11\% and 90.60\%, with a difference of
0.49 percentage points.
Figure~\ref{fig:distributions} shows the distribution of accuracy and net
confidence across radiologists in each assistance condition.

\begin{figure}[!htbp]
\centering
\includegraphics[width=\linewidth]{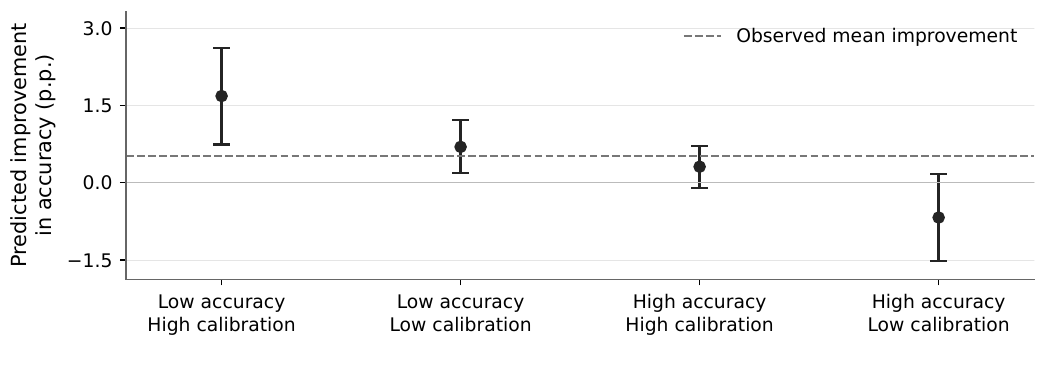}
\caption{Fitted improvement in accuracy with AI, following Figure~2 of
Caplin et al.~\cite{caplin2026abcs}. Low and high denote one standard
deviation below and above the mean. Predictions use the primary joint model,
averaging design indicators over the 8,500 observations. Bars are 95\%
confidence intervals using a standard normal approximation and standard
errors clustered by radiologist. The dashed line shows the observed mean
improvement of 0.52 percentage points.}
\Description{Four fitted profiles combine low or high accuracy with low or
high calibration. Low accuracy with high calibration has the largest fitted
improvement, and high accuracy with low calibration has the smallest.
Error bars show 95 percent confidence intervals.}
\label{fig:profiles}
\end{figure}

Table~\ref{tab:main} compares the accuracy association with and without
controlling for calibration. Without calibration, the accuracy coefficient is
$\AccuracyOnlyEstimate$ (SE \AccuracyOnlySE, $p=\AccuracyOnlyP$), which
does not reach significance at 5\%. Holding calibration fixed, a
one-standard-deviation increase in unassisted accuracy is associated with
a change of $\MainAbilityEstimate$ percentage points in improvement with AI
(SE \MainAbilitySE, $p=\MainAbilityP$), about twice the magnitude of the
accuracy-only coefficient. Holding accuracy fixed, a one-standard-deviation
increase in the calibration score is associated with an improvement that is
\MainCalibrationEstimate{} percentage points greater
(SE \MainCalibrationSE, $p=\MainCalibrationP$). Both conditional associations
are significant at 5\%: lower accuracy and a smaller absolute confidence gap
are associated with greater improvement. Supplementary
Table~\ref{supp-tab:predictors} reports the separate and joint models,
including the interaction specification.
Figure~\ref{fig:profiles} shows the joint model's predicted improvement at
low and high values of accuracy and calibration.

\begin{table}[!htbp]
\centering
\caption{Accuracy and improvement with AI, with and without calibration}
\label{tab:main}
\small
\setlength{\tabcolsep}{4pt}
\begin{tabular}{@{}p{0.35\linewidth}rrrr@{}}
\toprule
Specification & Accuracy (SE) & $p$ & Calibration (SE) & $p$ \\
\midrule
\multicolumn{5}{@{}l}{\textit{Designs~1--3: 227 radiologists, 8,500 observations}} \\
Without calibration & $-0.339$ (0.174) & 0.052 & --- & --- \\
With calibration & $-0.685$ (0.247) & 0.005 & $0.493$ (0.221) & 0.026 \\
\midrule
\multicolumn{5}{@{}l}{\textit{Designs~2--3: 68 radiologists, 3,730 observations}} \\
Without calibration & $-0.243$ (0.153) & 0.113 & --- & --- \\
With calibration & $-0.569$ (0.194) & 0.003 & $0.515$ (0.206) & 0.013 \\
 
\bottomrule
\end{tabular}
\begin{minipage}{\linewidth}
\footnotesize
\vspace{4pt}
All regressions include design fixed effects. The outcome is
improvement in accuracy with AI, measured in percentage points. Coefficients
give the association for one standard deviation in a predictor, using the
same standardization constants in both models within each sample.
Predictors are restandardized in the Designs~2--3 sample. Observations are
radiologist-patient pairs. The unassisted error rates are
\MainUnassistedErrorRate\% in Designs~1--3 and
\MainPairedUnassistedErrorRate\% in Designs~2--3. Parentheses contain standard errors.
P-values are two-sided and use a standard normal approximation with
standard errors clustered by radiologist. A dash indicates that calibration is omitted.
\end{minipage}
\end{table}

To assess whether the joint pattern is driven by the different method for
measuring improvement with AI assistance in Design~1, I restrict the sample
to Designs~2 and~3, which compare the same patient cases with and without AI.
The second panel of Table~\ref{tab:main} reports these results. Both joint
associations retain their signs and are statistically significant at 5\%.

Giving each radiologist equal total regression weight produces
coefficients of $-0.704$ for ability and 0.501 for calibration, close to
the primary point estimates. Supplementary Section~\ref{supp-sec:checks} reports this
comparison and the design-specific, clinical-history, reference, and
alternative-score analyses.

\FloatBarrier
\section{Discussion}

In the primary joint specification, lower unassisted accuracy and smaller
absolute confidence gaps are associated with greater improvements in accuracy with
AI. The calibration coefficient compares radiologists with similar
unassisted accuracy.

This conditional comparison connects Caplin et al.'s findings to
professional diagnostic judgments and complements Yu et al.'s focus on
unassisted performance. Within the primary specification, considering
accuracy and calibration together reveals an association that an
accuracy-only analysis misses.

\section*{Acknowledgments}
I thank Alex Moehring for helpful comments.
This work was supported by the Alfred P. Sloan Foundation through the
Cognitive Economics at Work grant.

\section*{Competing interests}
The author is a coauthor of the Caplin et al.\ paper examined here and declares no
other competing interests.

\section*{Data and code availability}
The source is the public Collab-CXR repository~\cite{moehring2025repository}.
The accompanying arXiv ancillary replication archive contains
the analysis input, source scripts, generated estimates, and instructions
for reproducing the analysis. Supplementary
Section~\ref{supp-sec:reproducibility} identifies the exact input by its
checksum and documents integrity checks. No new participant data were collected.

\section*{Use of generative AI}
OpenAI ChatGPT and Codex assisted with manuscript drafting and editing,
specification development, code implementation and debugging, and
interpretation of results. The author remains responsible for the final
analytical decisions and submitted content.

\clearpage
\bibliographystyle{ACM-Reference-Format}
\bibliography{references}


\begin{thebibliography}{5}


\ifx \showCODEN    \undefined \def \showCODEN     #1{\unskip}     \fi
\ifx \showDOI      \undefined \def \showDOI       #1{#1}\fi
\ifx \showISBNx    \undefined \def \showISBNx     #1{\unskip}     \fi
\ifx \showISBNxiii \undefined \def \showISBNxiii  #1{\unskip}     \fi
\ifx \showISSN     \undefined \def \showISSN      #1{\unskip}     \fi
\ifx \showLCCN     \undefined \def \showLCCN      #1{\unskip}     \fi
\ifx \shownote     \undefined \def \shownote      #1{#1}          \fi
\ifx \showarticletitle \undefined \def \showarticletitle #1{#1}   \fi
\ifx \showURL      \undefined \def \showURL       {\relax}        \fi
\providecommand\bibfield[2]{#2}
\providecommand\bibinfo[2]{#2}
\providecommand\natexlab[1]{#1}
\providecommand\showeprint[2][]{arXiv:#2}

\bibitem[Agarwal et~al\mbox{.}(2023)]%
        {agarwal2023radiology}
\bibfield{author}{\bibinfo{person}{Nikhil Agarwal}, \bibinfo{person}{Alex
  Moehring}, \bibinfo{person}{Pranav Rajpurkar}, {and} \bibinfo{person}{Tobias
  Salz}.} \bibinfo{year}{2023}\natexlab{}.
\newblock \bibinfo{booktitle}{\emph{Combining human expertise with artificial
  intelligence: Experimental evidence from radiology}}.
\newblock \bibinfo{type}{NBER Working Paper} 31422.
  \bibinfo{institution}{National Bureau of Economic Research}.
\newblock
\urldef\tempurl%
\url{https://doi.org/10.3386/w31422}
\showDOI{\tempurl}


\bibitem[Caplin et~al\mbox{.}(2026)]%
        {caplin2026abcs}
\bibfield{author}{\bibinfo{person}{Andrew Caplin}, \bibinfo{person}{David~J.
  Deming}, \bibinfo{person}{Shangwen Li}, \bibinfo{person}{Daniel Martin},
  \bibinfo{person}{Philip Marx}, \bibinfo{person}{Ben Weidmann}, {and}
  \bibinfo{person}{Kadachi~Jiada Ye}.} \bibinfo{year}{2026}\natexlab{}.
\newblock \showarticletitle{The {ABCs} of who benefits from working with {AI}:
  Ability, beliefs, and calibration}.
\newblock \bibinfo{journal}{\emph{Management Science}} \bibinfo{volume}{72},
  \bibinfo{number}{7} (\bibinfo{year}{2026}), \bibinfo{pages}{5843--5852}.
\newblock
\urldef\tempurl%
\url{https://doi.org/10.1287/mnsc.2024.08994}
\showDOI{\tempurl}


\bibitem[Moehring(2025)]%
        {moehring2025repository}
\bibfield{author}{\bibinfo{person}{Alex Moehring}.}
  \bibinfo{year}{2025}\natexlab{}.
\newblock \bibinfo{title}{{Collab-CXR}: A dataset for understanding
  radiologist-artificial intelligence collaboration}.
\newblock \bibinfo{howpublished}{Open Science Framework}.
\newblock
\urldef\tempurl%
\url{https://doi.org/10.17605/OSF.IO/Z7APQ}
\showDOI{\tempurl}
\newblock
\shownote{Repository metadata updated September 29, 2025}.


\bibitem[Moehring et~al\mbox{.}(2025)]%
        {moehring2025dataset}
\bibfield{author}{\bibinfo{person}{Alex Moehring}, \bibinfo{person}{Manasi
  Kutwal}, \bibinfo{person}{Ray Huang}, \bibinfo{person}{Oishi Banerjee},
  \bibinfo{person}{Adam Jacobi}, \bibinfo{person}{Corey Eber},
  \bibinfo{person}{Dexter Mendoza}, \bibinfo{person}{Mike Chung},
  \bibinfo{person}{Etan Dayan}, \bibinfo{person}{Yogesh Gupta},
  \bibinfo{person}{Tan D.~T. Bui}, \bibinfo{person}{Steven Q.~H. Truong},
  \bibinfo{person}{Anuj Pareek}, \bibinfo{person}{Curtis~P. Langlotz},
  \bibinfo{person}{Matthew~P. Lungren}, \bibinfo{person}{Nikhil Agarwal},
  \bibinfo{person}{Pranav Rajpurkar}, {and} \bibinfo{person}{Tobias Salz}.}
  \bibinfo{year}{2025}\natexlab{}.
\newblock \showarticletitle{A dataset for understanding radiologist-artificial
  intelligence collaboration}.
\newblock \bibinfo{journal}{\emph{Scientific Data}}  \bibinfo{volume}{12},
  Article \bibinfo{articleno}{739} (\bibinfo{year}{2025}),
  \bibinfo{numpages}{7}~pages.
\newblock
\urldef\tempurl%
\url{https://doi.org/10.1038/s41597-025-05054-0}
\showDOI{\tempurl}


\bibitem[Yu et~al\mbox{.}(2024)]%
        {yu2024heterogeneity}
\bibfield{author}{\bibinfo{person}{Feiyang Yu}, \bibinfo{person}{Alex
  Moehring}, \bibinfo{person}{Oishi Banerjee}, \bibinfo{person}{Tobias Salz},
  \bibinfo{person}{Nikhil Agarwal}, {and} \bibinfo{person}{Pranav Rajpurkar}.}
  \bibinfo{year}{2024}\natexlab{}.
\newblock \showarticletitle{Heterogeneity and predictors of the effects of {AI}
  assistance on radiologists}.
\newblock \bibinfo{journal}{\emph{Nature Medicine}} \bibinfo{volume}{30},
  \bibinfo{number}{3} (\bibinfo{year}{2024}), \bibinfo{pages}{837--849}.
\newblock
\urldef\tempurl%
\url{https://doi.org/10.1038/s41591-024-02850-w}
\showDOI{\tempurl}


\end{thebibliography}

\clearpage
\setcounter{section}{0}
\setcounter{table}{0}
\renewcommand{\thesection}{S\arabic{section}}
\renewcommand{\thetable}{S\arabic{table}}
\renewcommand{\theHsection}{supp.\arabic{section}}
\renewcommand{\theHtable}{supp.\arabic{table}}
\setlength{\parskip}{4pt}
\setlength{\parindent}{0pt}
\begin{center}
  {\Large Supplementary material}
\end{center}
\section{Sample and measurement}
\label{supp-sec:sample}

\subsection*{Source and sample flow}
The source is the public Collab-CXR repository~\cite{moehring2025repository},
described by Moehring et al.~\cite{moehring2025dataset}. The input is
\texttt{data\_public.txt.gz}, identified by its checksum in
Section~\ref{supp-sec:reproducibility}. The primary diagnostic reference is the
patient-pathology mean probability among the five expert-panel
radiologists identified in the source file. It is fixed across the
experimental radiologist's conditions. Probabilities in the public file
are encoded on the unit interval. The analysis helpers use an
internal percentage scale for some calculations, with the same
threshold classifications and percentage-point outcome.

\begin{table}[ht]
\centering
\caption{Sample construction}
\label{supp-tab:sample_flow}
\begin{tabular}{@{}lrrr@{}}
\toprule
Stage & Assessment rows & Radiologists & Patients\\
\midrule
Public file & 2,717,728 & 237 & 324 \\
Designs 1--3 & 2,380,768 & 227 & 324 \\
Exclude warm-up observations & 2,179,840 & 227 & 324 \\
Exclude pilot identifiers & 2,179,840 & 227 & 324 \\
Keep 15 tasks with AI predictions & 314,400 & 227 & 324 \\
Require complete analysis fields & 314,400 & 227 & 324 \\
 
\bottomrule
\end{tabular}
\end{table}

The pilot filter removes no additional rows after the preceding
restrictions in Table~\ref{supp-tab:sample_flow}. The 15 retained tasks are abnormal, airspace opacity,
atelectasis, bacterial/lobar pneumonia, cardiomediastinal abnormality,
cardiomegaly, consolidation, edema, lesion, pleural effusion, pleural
other, pneumothorax, rib fracture, shoulder fracture, and support
device/hardware.

The regression sample contains 8,500 radiologist-patient observations:
4,770 from 159 Design 1 radiologists, 1,980 from 33 Design 2
radiologists, and 1,750 from 35 Design 3 radiologists.
Design 1 uses 30 outcome cases assessed without AI per radiologist and a separate
30-patient assisted average. No radiologist's assisted and unassisted
Design 1 sets overlap. Assisted sets differ across radiologists and overlap
across radiologists. Designs 2 and 3 contribute 60 and 50 patients per radiologist,
respectively. Where both clinical-history states are available for the
same radiologist, case, and AI state, they are equally weighted.

The supplied analysis code for Yu et al.~\cite{yu2024heterogeneity}
excludes participants assigned to sequential AI presentation and retains
those assigned to simultaneous presentation. This restriction is established from that code,
not an explicit statement in the published article. The public treatment
field does not identify this participant-level assignment, and case order
and session fields do not recover it. The primary analysis therefore
includes all eligible radiologists in Designs 1--3. For one sensitivity
analysis, I use the supplied
\texttt{radiologist\_ids\_for\_replication.txt} list to define a
140-radiologist subsample. Its source and matching
procedure are not independently documented, so I do not treat it as a
verified reconstruction of Yu et al.'s sample.

\subsection*{Patient omission and standardization}
Let $b\in[0,1]$ denote a recorded radiologist probability,
$c=\max\{b,1-b\}$ its decision confidence, and $a$ binary agreement with
the expert-panel reference label. Both the radiologist's decision and the
reference label are positive only when their probabilities exceed 0.5.

Caplin et al.~\cite{caplin2026abcs} compute ability as mean accuracy and calibration as the negative
absolute gap between mean confidence and accuracy, both from unassisted
probability assessments for individual images. The original regression uses one
observation per participant, relating accuracy on a separate image block
to randomized AI access, both traits, and their interactions with AI
access. Here the outcome is improvement with AI, and trait construction
excludes each patient case in turn.

For radiologist $h$ and patient case $i$, I construct both traits from the
radiologist's unassisted assessments of all other patient cases. Both history states
are retained when available. Standardization first averages each trait
across patient cases within radiologist. It then computes the unweighted
mean and sample standard deviation (denominator $H-1$) across those
radiologist means, where $H$ is the number of radiologists in the estimation
sample, and applies the resulting constants to each regression
observation. Thus the observation-level standardized values need not
have exactly zero mean and unit variance.

Primary regression weights are equal across radiologist-patient
observations. Equal total weights for radiologists instead make a radiologist's total
contribution independent of the number of patient cases contributing regression observations. In
clinical-history sensitivities, traits and outcomes are constructed
separately within each history stratum. Each history sample contains 5,240
observations, and the samples overlap because Design 2 radiologist-patient
pairs can appear in both.

\section{Estimates and sensitivity analysis}
\label{supp-sec:checks}

Table~\ref{supp-tab:checks} reports the primary estimate and exploratory
specification and measurement checks covering designs, weighting,
confidence definitions, and reference construction.
Table~\ref{supp-tab:ci} gives their 95\% confidence intervals based on a standard
normal approximation, using the same samples as Table~\ref{supp-tab:checks}.

\begingroup
\small
\setlength{\tabcolsep}{3pt}
\begin{longtable}{@{}>{\raggedright\arraybackslash}p{0.29\linewidth}rr rrrr@{}}
\caption{Specification and measurement checks}\label{supp-tab:checks}\\
\toprule
Specification & Radiologists & Obs. & Accuracy (SE) & $p$ & Score (SE) & $p$\\
\midrule
\endfirsthead
\multicolumn{7}{l}{Table \thetable\ continued}\\
\toprule
Specification & Radiologists & Obs. & Accuracy (SE) & $p$ & Score (SE) & $p$\\
\midrule
\endhead
\bottomrule
\endfoot
All designs (primary) & 227 & 8,500 & $-0.685$ (0.247) & 0.005 & $0.493$ (0.221) & 0.026 \\
Designs 1--2 & 192 & 6,750 & $-0.654$ (0.293) & 0.026 & $0.393$ (0.261) & 0.132 \\
Designs 2--3 & 68 & 3,730 & $-0.569$ (0.194) & 0.003 & $0.515$ (0.206) & 0.013 \\
Design 1 & 159 & 4,770 & $-0.748$ (0.399) & 0.061 & $0.481$ (0.354) & 0.174 \\
Design 2 & 33 & 1,980 & $-0.412$ (0.142) & 0.004 & $0.153$ (0.149) & 0.302 \\
Design 3 & 35 & 1,750 & $-0.697$ (0.359) & 0.052 & $0.812$ (0.312) & 0.009 \\
Supplied 140-radiologist subsample & 140 & 5,190 & $-0.431$ (0.333) & 0.195 & $0.337$ (0.290) & 0.246 \\
No design fixed effects & 227 & 8,500 & $-0.635$ (0.235) & 0.007 & $0.400$ (0.201) & 0.047 \\
Equal total radiologist weight & 227 & 8,500 & $-0.704$ (0.289) & 0.015 & $0.501$ (0.260) & 0.054 \\
Design and patient fixed effects & 227 & 8,500 & $-0.761$ (0.227) & 0.001 & $0.559$ (0.206) & 0.007 \\
Net confidence & 227 & 8,500 & $-0.579$ (0.248) & 0.020 & $-0.343$ (0.222) & 0.121 \\
Negative ten-bin ECE & 227 & 8,500 & $-0.655$ (0.301) & 0.029 & $0.363$ (0.307) & 0.238 \\
Without clinical history & 227 & 5,240 & $-0.908$ (0.348) & 0.009 & $0.636$ (0.342) & 0.063 \\
With clinical history & 227 & 5,240 & $-0.509$ (0.237) & 0.032 & $0.263$ (0.224) & 0.239 \\
Stable public-data peer reference & 227 & 8,500 & $-0.746$ (0.238) & 0.002 & $0.671$ (0.202) & 0.001 \\
Median of five-radiologist expert panel & 227 & 8,500 & $-0.484$ (0.229) & 0.035 & $0.406$ (0.194) & 0.037 \\
 
\end{longtable}
\endgroup

Coefficients in Tables~\ref{supp-tab:checks} and~\ref{supp-tab:ci} are percentage-point associations per standard
deviation of the predictor. The second predictor is the primary
negative absolute net confidence score except in the two
alternative-score rows. Accuracy and the score enter jointly. Design
indicators are included when applicable, except in the explicitly
labeled no-design-effects row. Predictors are restandardized within each
sample using the mean and sample standard deviation of their
radiologist-level means. Parentheses contain standard errors.
P-values are two-sided and use a standard normal approximation with
standard errors clustered by radiologist. Rounding to
three decimals can display a p-value just below 0.001 as 0.001.

\subsection*{Outcome scale}
The main coefficients remain in percentage points. To express them in
outcome standard deviations, I first calculate improvement in accuracy
with AI, then subtract its mean and divide by its sample
standard deviation of \MainDifferenceSD{} percentage points across the
8,500 equally weighted regression observations. I do not standardize
assisted and unassisted accuracy separately. With the same predictors
and design indicators, the coefficients are
$\StandardizedOutcomeAbilityEstimate$ (SE \StandardizedOutcomeAbilitySE,
$p=\MainAbilityP$) for ability and
$\StandardizedOutcomeCalibrationEstimate$ (SE \StandardizedOutcomeCalibrationSE,
$p=\MainCalibrationP$) for calibration. They describe differences in
improvement with AI per standard deviation of the respective trait. Rescaling
the outcome changes the units, not these slopes' p-values.

\subsection*{Distinct confidence constructs}
Here $\bar c$ and $\bar a$ denote average decision confidence and average
binary agreement over the focal radiologist's recorded unassisted
assessments, excluding patient case $i$.
Net confidence is $\bar c-\bar a$, rather than its negative
absolute value. Negative ten-bin expected calibration error is minus
the frequency-weighted absolute difference between the mean recorded
probability and the fraction of positive reference labels within
probability bins. With probability $b\in[0,1]$ and reference label $s$,
the bin is $\min\{\lfloor 10b\rfloor,9\}$, including $b=1$ in the final
bin. Unlike the primary score, this compares probabilities of pathology
presence with reference-label frequencies, rather than decision
confidence with decision accuracy. As in the primary specification, these scores use
unassisted assessments excluding patient case $i$ and are standardized using
radiologist means.

\subsection*{Alternative reference labels}
The panel-median sensitivity replaces the mean of the five expert
probabilities with their median. I construct a peer reference that remains
fixed across the focal radiologist's conditions because the supplied
\texttt{gt\_average\_experiment} field can produce different reference labels
across AI states. This peer reference
instead averages the experimental radiologists' unassisted assessments with
clinical history. The focal radiologist is excluded if that radiologist
contributed to the relevant patient-pathology pool. The resulting
reference is then fixed across that radiologist's AI and history conditions.
At least six other radiologists contribute to every reference. Its
probabilities range from 0 to 0.98865. The replication script asserts
probability bounds and invariance within each focal
radiologist-patient-pathology cell.

\subsection*{Inference and interpretation}
For the estimates tabulated in the main letter and in Tables~\ref{supp-tab:checks},
\ref{supp-tab:ci}, and~\ref{supp-tab:predictors}, p-values are two-sided and use a
standard normal approximation with standard errors clustered by radiologist.
Confidence intervals use the same approximation. Design-specific rows have
fewer radiologist clusters than the primary sample of 227 radiologists. Patient fixed
effects address patient-level mean heterogeneity and do not substitute
for a case-dependence covariance adjustment. Patient omission separates
outcomes from trait measurement, but observations can still share assisted
cases in Design 1.

For the primary specification, two-way clustering by radiologist and
patient case gives standard errors of \TwoWayAbilitySE{} and \TwoWayCalibrationSE{}.
A radiologist-pairs bootstrap with \BootstrapReps{} draws and random seed
\BootstrapSeed{} yields 95\% percentile-bootstrap confidence intervals
$[\BootstrapAbilityLow,\BootstrapAbilityHigh]$ and
$[\BootstrapCalibrationLow,\BootstrapCalibrationHigh]$.
The bootstrap resamples radiologists with replacement while holding the
constructed observations and standardization constants fixed.
Radiologist resampling conditions on the
observed cases. Neither that bootstrap nor the two-way calculation
fully represents dependence through all patients in the
radiologist-specific Design 1 assisted averages. The letter therefore
limits its inferential interpretation to the observed case data.

There are \ReferenceProbabilityTies{} expert-panel mean references and
\ReaderProbabilityTies{} retained radiologist assessments exactly equal to
0.5. The main convention treats a report of 0.5 as negative, whereas
Caplin et al.~\cite{caplin2026abcs} assign half credit for such a report.
Classifying these radiologist reports as positive and reconstructing the
traits and outcomes changes the ability coefficient by
$\TieAbilityChange$ and the calibration coefficient by
$\TieCalibrationChange$. The accuracy coefficient remains negative and the
calibration coefficient remains positive.

\begingroup
\small
\setlength{\tabcolsep}{4pt}
\begin{longtable}{@{}>{\raggedright\arraybackslash}p{0.35\linewidth}rr@{}}
\caption{95\% confidence intervals based on a standard normal approximation}\label{supp-tab:ci}\\
\toprule
Specification & Accuracy [95\% CI] & Score [95\% CI]\\
\midrule
\endfirsthead
\multicolumn{3}{l}{Table \thetable\ continued}\\
\toprule
Specification & Accuracy [95\% CI] & Score [95\% CI]\\
\midrule
\endhead
\bottomrule
\endfoot
All designs (primary) & $-0.685$ [-1.168, -0.202] & $0.493$ [0.060, 0.926] \\
Designs 1--2 & $-0.654$ [-1.229, -0.079] & $0.393$ [-0.119, 0.904] \\
Designs 2--3 & $-0.569$ [-0.949, -0.189] & $0.515$ [0.110, 0.919] \\
Design 1 & $-0.748$ [-1.530, 0.034] & $0.481$ [-0.212, 1.174] \\
Design 2 & $-0.412$ [-0.690, -0.133] & $0.153$ [-0.138, 0.444] \\
Design 3 & $-0.697$ [-1.400, 0.007] & $0.812$ [0.200, 1.424] \\
Supplied 140-radiologist subsample & $-0.431$ [-1.083, 0.221] & $0.337$ [-0.232, 0.905] \\
No design fixed effects & $-0.635$ [-1.096, -0.175] & $0.400$ [0.006, 0.794] \\
Equal total radiologist weight & $-0.704$ [-1.271, -0.138] & $0.501$ [-0.009, 1.011] \\
Design and patient fixed effects & $-0.761$ [-1.207, -0.315] & $0.559$ [0.155, 0.962] \\
Net confidence & $-0.579$ [-1.064, -0.093] & $-0.343$ [-0.778, 0.091] \\
Negative ten-bin ECE & $-0.655$ [-1.245, -0.065] & $0.363$ [-0.240, 0.965] \\
Without clinical history & $-0.908$ [-1.590, -0.227] & $0.636$ [-0.035, 1.308] \\
With clinical history & $-0.509$ [-0.974, -0.044] & $0.263$ [-0.175, 0.702] \\
Stable public-data peer reference & $-0.746$ [-1.213, -0.279] & $0.671$ [0.275, 1.066] \\
Median of five-radiologist expert panel & $-0.484$ [-0.933, -0.035] & $0.406$ [0.025, 0.787] \\
 
\end{longtable}
\endgroup

\subsection*{Joint and separate predictors}
Does the association with accuracy change when calibration is included?
Table~\ref{supp-tab:predictors} holds the sample of 227 radiologists, 8,500 observations,
standardization constants, and design indicators fixed while varying
the predictors. Accuracy alone has a coefficient of
$\AccuracyOnlyEstimate$ (SE \AccuracyOnlySE, $p=\AccuracyOnlyP$), which
is not significant at 5\%. The joint model retains the negative sign,
with a larger coefficient in absolute value. Since the sample and
standardization are held fixed, this comparison isolates the change in
the accuracy association when calibration is included.

The calibration-only point estimate is close to zero. Calibration's positive
coefficient in the joint model describes a conditional
association among radiologists with similar unassisted accuracy.
The interaction estimate provides no clear evidence that this calibration
association varies with accuracy. In that model, the accuracy and calibration
coefficients describe the association when the other standardized predictor is zero.

\begin{table}[htbp]
\centering
\caption{Joint and separate predictors of improvement in accuracy with AI}
\label{supp-tab:predictors}
\small
\setlength{\tabcolsep}{3pt}
\begin{tabular}{@{}lrrrrrr@{}}
\toprule
Specification & Accuracy (SE) & $p$ & Calibration (SE) & $p$ & Interaction (SE) & $p$\\
\midrule
Accuracy only & $-0.339$ (0.174) & 0.052 & --- & --- & --- & --- \\
Calibration only & --- & --- & $-0.009$ (0.171) & 0.960 & --- & --- \\
Accuracy and calibration & $-0.685$ (0.247) & 0.005 & $0.493$ (0.221) & 0.026 & --- & --- \\
With interaction & $-0.715$ (0.252) & 0.005 & $0.458$ (0.214) & 0.033 & $-0.082$ (0.128) & 0.523 \\
 
\bottomrule
\end{tabular}
\begin{minipage}{\linewidth}
\footnotesize
\vspace{4pt}
The outcome is improvement in accuracy with AI, measured in percentage
points relative to accuracy without AI. All models include design indicators. Both traits use
the primary 15-task measures and the same full-sample standardization.
The interaction is the product of standardized accuracy and calibration.
P-values are two-sided and use a standard normal approximation with
standard errors clustered by radiologist. The ``Accuracy and calibration''
row is the primary model.
\end{minipage}
\end{table}

\FloatBarrier
\section{Reproducibility and disclosures}
\label{supp-sec:reproducibility}
The accompanying replication package is self-contained, with the unchanged
input, analysis helpers, a runner, and generated numerical and TeX outputs.
The repository's current citation metadata list Alex Moehring as creator,
2025 as the publication year, and September 29, 2025 as the update date.
The dataset article cites a 2023 record updated April 3, 2025. These
repository dates do not establish the version of an individual data file.
The input's SHA-256 is
\begin{center}
\footnotesize
\texttt{dde3c7391a21b3c146b29edd5b346e86ef80e8393548cbbd2b142039815d5e03}.
\end{center}
Run \texttt{python3 -B replication/rebuild\_ailet.py} from the project
directory with NumPy, pandas, and Matplotlib installed. Add
\texttt{--verify-original} to compare all 8,500 primary regression
points against a patient-by-patient implementation.
The two implementations differ by at most
$7.11\times10^{-15}$. The script also checks the primary estimates,
sample counts and the stable peer reference.
The package README records software requirements and each output.
The data remain subject to the source repository's license.

The runner also computes the two-way clustering, 999-draw bootstrap,
and exact-threshold checks reported in Section~\ref{supp-sec:checks} and tests
numerical agreement with reference outputs. It generates all three count
columns in Table~\ref{supp-tab:sample_flow} and the manuscript values for these
checks. A coverage index maps the reported analyses to their code and
output files. The package README documents the execution steps and
validation checks.
The runner also generates the letter's two figures and saves their
radiologist summaries, histogram counts, fitted profiles, and full coefficient
covariance matrix. The fitted profiles use linear combinations of the primary
regression coefficients with design indicators averaged over observations.

The primary analysis is a secondary analysis of a public experiment
and was not preregistered. The author is a coauthor of the
Caplin et al.\ paper. OpenAI ChatGPT and Codex assisted with drafting, editing,
specification development, implementation, debugging, and
interpretation. The author remains responsible for final analytical
decisions and submitted content.

\end{document}